\documentclass[prb,aps,amssymb,showpacs,twocolumn]{revtex4}
\usepackage{graphicx,amssymb,amsmath,color,psfrag}
\usepackage{amsthm}
\usepackage{amsfonts}
\usepackage{algorithmic}
\usepackage{enumerate}
\usepackage{latexsym}

\begin{document}

\title{NMR relaxation rate and static spin susceptibility in graphene}
\author{Tianxing Ma, Bal\'azs D\'ora}
 \affiliation{Max-Planck-Institut f\"ur Physik Komplexer Systeme, N\"othnitzer Str. 38, 01187 Dresden, Germany}
\date{\today}

\begin{abstract}
The NMR relaxation rate and the static spin susceptibility in
graphene are studied within a tight-binding description. At half
filling, the NMR relaxation rate follows a power law as $T^2$ on the
particle-hole symmetric side, while with a finite chemical potential
$\mu$ and next-nearest neighbor $t'$, the $(\mu+3t')^2$ terms
dominate at low excess charge $\delta$. The static spin
susceptibility is linearly dependent on temperature $T$ at half
filling when $t'=0$, while with a finite $\mu$ and $t'$, it should
be dominated by $(\mu+3t')$ terms in low energy regime. These
unusual phenomena are direct results of the low energy excitations
of graphene, which behave as massless Dirac fermions. Furthermore,
when $\delta$ is high enough, there is a pronounced crossover which
divides the temperature dependence of the NMR relaxation rate and
the static spin susceptibility into two temperature regimes: the NMR
relaxation rate and the static spin susceptibility increase
dramatically as temperature increases in the low temperature regime,
and after the crossover, both decrease as temperature increases at
high temperatures. This crossover is due to the well-known
logarithmic Van Hove singularity in the density of states, and its
position dependence of temperature is sensitive to $\delta$.
\end{abstract}


\maketitle
\section{Introduction}

Graphene, the latest carbon allotrope to be
discovered\cite{Novoselov}, is made out of carbon atoms organized
into a honeycomb lattice. The characteristics of the honeycomb
lattice make graphene a half filled system with a density of states
(DOS) that vanishes linearly at the neutrality point, and an
effective, low energy quasiparticle spectrum characterized by a
dispersion which is linear in momentum close to the Fermi
energy\cite{Wallace}. These two features underlie the unconventional
electronic properties of this material, whose low energy excitations
behave as massless Dirac fermions\cite{Semenof,Geim}.

There have been intensive theoretical and experimental studies on
graphene to this date, for instance, half integer and unconventional
quantum Hall effect\cite{Novoselov1,Yuanbo,Novoselov2,Yang}, quantum
minimum conductivity\cite{Novoselov1,Ziegler,MacDonald}, bipolar
supercurrent\cite{Heersche}, ferromagnetism\cite{Vozmediano},
optical conductivity\cite{Jiang} and the possibility of
superconductivity\cite{Super}. On the other hand, the most
interesting and promising properties from the technological point of
view are its great crystalline quality, high mobility and resilience
to very high current densities\cite{Geim}, the ability to tune the
carrier density through a gate voltage\cite{Novoselov}, as well as
the fact that graphene exhibits both spin and valley degrees of
freedom which might be harnessed in envisaged
spintronics\cite{Kane,Cho,Qi}. For a review of other remarkable
properties of such systems as well as a discussion of possible
technological applications, the reader is referred to
Ref.\cite{Geim,Rmp}. Graphene is poised to become a new paradigm in
solid state physics and materials science.

Nuclear magnetic resonance (NMR) is usually an excellent technique
for probing the electronic properties of materials as it is
sensitive to the DOS near the Fermi edge, and this method allows one
to study static magnetic correlations and low-energy spin
excitations. For a material with a Fermi liquid state, the
temperature dependent spin-lattice relaxation time $T_{1}$ follows
the well-known Korringa relation where $1/T_{1}$ varies linearly
with temperature\cite{NMR}. As one of the most powerful methods for
investigating mechanisms of superconductivity of the many exotic
types of superconductors being discovered today, it has turned out
that the explanation for the peak in the NMR relaxation rate just in
superconducting state was that the DOS peaked dramatically at the
edge of a energy gap\cite{NMR,Peak}. Recent NMR experiments by
Singer \textit{et al.} showed a deviation from Fermi liquid behavior
in carbon nanotubes with an energy gap evident at low
temperatures\cite{Singer}. In the framework of the
Tomonaga-Luttinger liquid, the low temperature properties are
governed by a gapped relaxation due to a spin gap, which crosses
over smoothly to the Luttinger liquid behavior with increasing
temperature\cite{nmrnt}.

In spite of being few atoms thick, the system of graphene was found
to be stable and ready for exploration\cite{Geim}, and, it is
believed that NMR should provide cornucopian and significant
information on the electronic properties of graphene\cite{NMR}. In
the present paper, we study the NMR relaxation rate and the static
spin susceptibility of graphene within a tight-binding description.
At half filling, the NMR relaxation rate follows a power law as
$T^2$ on the particle-hole symmetric side, while away from half
filling and with a finite next-nearest neighbor $t'$, the
$(\mu+3t')^2$ terms dominate at low excess charge $\delta$. The
static spin susceptibility is linearly dependent on temperature $T$
at half filling when $t'=0$, while with a finite $\mu$ and $t'$, it
should be dominated by $(\mu+3t')$ terms in low energy regime. These
unusual phenomena are direct results of the low energy excitations
of graphene, which behave as massless Dirac fermions. Furthermore,
when $\delta$ is high enough, there is a pronounced crossover which
divides the temperature dependence of the NMR relaxation rate and
the static spin susceptibility into two temperature regimes: the NMR
relaxation rate and the static spin susceptibility increase
dramatically as temperature increases in the low temperature regime,
and after the crossover, both decrease as temperature increases at
high temperatures. This crossover is due to the well-known
logarithmic Van Hove singularity in the DOS, and its position
dependence of temperature is sensitive to $\delta$.

The rest of the paper is organized as follows. The theoretical
framework is introduced in section II. Our numerical result and
discussion are shown in section III, and the paper is concluded with
a summary in section IV.

\section{Theoretical framework}
Graphene is a two dimensional crystal of carbon atoms with a
honeycomb lattice, which can be described in terms of two
interpenetrating triangular sublattices, A and B, and then the
electronic structure of graphene can be captured within a
tight-binding description\cite{Rmp,Model,Dos,tp}
\begin{eqnarray}
H&=&-t\sum_{i\eta\sigma}(a_{i\sigma}^{\dagger}
b_{i+\eta\sigma})+t'\sum_{i\gamma\sigma}
(a_{i\sigma}^{\dagger}a_{i+\gamma\sigma}+b_{i\sigma}^{\dagger}b_{i+\gamma\sigma})+{\rm
h.c.}\nonumber \\
&+&\mu\sum_{i\sigma} (a_{i\sigma }^{\dagger }a_{i\sigma }+b_{i\sigma
}^{\dagger }b_{i\sigma }),
\end{eqnarray}
where $a_{i,\sigma}$ ($a_{i,\sigma}^{\dag}$) annihilates (creates)
electrons at the site ${\bf R}_i$ with spin $\sigma$
($\sigma=\uparrow,\downarrow$) on sublattice A, and $b_{i,\sigma}$
($b_{i,\sigma}^{\dag}$) annihilates (creates) electrons at the site
${\bf R}_i$ with spin $\sigma$ ($\sigma=\uparrow,\downarrow$) on
sublattice B. $t$ and $t'$ are the nearest neighbor and next-nearest
neighbor hopping energies respectively, and $\mu$ is the chemical
potential.  The presence of $t'$ introduces an asymmetry between the
valance and conduction bands, thus violating particle-hole symmetry.
Specific values for $t$ and $t'$ have been estimated\cite{tp} by
comparing a tight-binding description to first-principle
calculations. Following their estimates, we take $t=$2.7eV. To learn
more on the effect of $t'$ in graphene, cases with different values
of $t'$ will be studied, and a typical $t'=$0.27 eV will be paid
more attention\cite{Model} in the following.

In the sublattice system, there are two coupled sublattices, and the
energy spectrum has two branches. In this case, the one-particle
Green's functions are matrices
\begin{eqnarray}
g(i-j,\tau)=\left(
\begin{array}{cc}
g_{aa}(i-j,\tau)&g_{ab}(i-j,\tau)\\
g_{ba}(i-j,\tau)&g_{bb}(i-j,\tau)%
\end{array}
\right),
\end{eqnarray}
where the longitudinal and transverse parts are defined as
\begin{eqnarray}
g_{mn}(i-j,\tau)=-\langle T_{\tau}m_{i}(\tau)
n_{j}^{\dagger}(0)\rangle, m,n=a,b;
\end{eqnarray}with $\tau$ is the imaginary time, and $T_{\tau}$ is
the $\tau$ order operator. Then the Green's functions are obtained
as
\begin{eqnarray}
g_{aa}(k,\omega)={1\over 2}\sum_{\nu=1,2}{1\over \omega-
\xi^{(\nu)}_{k}}=g_{bb}(k,\omega),\nonumber \\
g_{ab}(k,\omega)={1\over
2}\frac{\phi_{k}}{|\phi_{k}|}\sum_{\nu=1,2}(-1)^{\nu} {1\over
\omega-\xi^{(\nu)}_{k}}=g^{*}_{ba}(k,\omega)
\end{eqnarray}
respectively, where
\begin{eqnarray}
\xi^{(\nu)}_{k}&=&t'\gamma_{k}+\mu+ 2t|\phi_{k}|(-1)^{\nu+1},\nonumber \\
\phi_{k}&=&[e^{ik_{x}}+e^{i(\frac{1}{2}k_{x}+\frac{\sqrt{3}}{2}k_{y})}+e^{i(\frac{1}{2}k_{x}-\frac{\sqrt{3}}{2}k_{y})}],\nonumber \\
\gamma_{k}&=&2[\cos{\sqrt{3}k_{x}}+2\cos{\frac{3}{2}k_{x}}\cos{(\frac{{\sqrt
3}}{2}k_{y})}].
\end{eqnarray}

From these, the DOS follows as
\begin{eqnarray}
\rho(\omega)=-\frac{2}{\pi}{1\over
N}\sum_{k}\textmd{Im}g_{aa}(k,\omega+i\Gamma),
\end{eqnarray}
with $\Gamma\rightarrow 0^+$. For $t'=0$ an analytical expression
for the DOS per unit cell can be derived\cite{Rmp}:
\begin{eqnarray}\label{Dostj}
\rho_{0}(\omega)&=&{1\over N}\sum_{k\nu}\delta[\omega-(-1)^{\nu}
t\sqrt{3+\gamma_{k}}]
\nonumber \\
&=&\displaystyle{\frac{2}{\pi^2} \frac{|\omega|}{t^2}
\frac{1}{\sqrt{Z_0}} \mathbf{K}[Z(\omega)]},
\end{eqnarray}
where $Z(\omega)=\sqrt{\frac{Z_1}{Z_0}}$ with
\begin{equation}\label{Dostj}
\begin{array}{l}
Z_{0}=\frac{1}{4}(1+\frac{|\omega|}{t})^{3}(3-\frac{|\omega|}{t}),
Z_{1}=4\frac{|\omega|}{t},  |\omega| \!\leq\! t \\ \\
Z_{1}=\frac{1}{4}(1+\frac{|\omega|}{t})^{3}(3-\frac{|\omega|}{t}),
Z_{0}=4\frac{|\omega|}{t},  t\!<\! |\omega |\!<\! 3t,
\end{array}
\end{equation}
and $\mathbf{K}(x)$ is the complete elliptic integral of the first
kind. With a finite $t'$, the DOS is evaluated by inserting unity in
the form of an integral over the Dirac delta function as
\begin{eqnarray}
\rho(\omega)&=&\frac 1N\sum_{k\nu}\delta[1-\xi(k)^{(\nu)}] \nonumber \\
&=&\frac 1N\sum_{k,\nu}\delta(1-\xi(k)^{(\nu)})\nonumber \\
&\times& \frac 12\sum_{\lambda=1,2}\int\limits_{-\infty}^\infty
d\epsilon \delta[\epsilon-(-1)^\lambda t\sqrt{3+\gamma_k}].
\end{eqnarray}
Then, by interchanging the integration with $k$ summation, we obtain
\begin{equation}
\rho(\omega)=\frac
{1}{2N}\sum_{k\nu\lambda}\int\limits_{-\infty}^\infty d\epsilon
\delta[\omega-t'(g_k^2-3)-(-1)^\nu t
g_k]\delta[\epsilon-(-1)^\lambda t g_k]
\end{equation}
with $g_k=\sqrt{3+\gamma_k}$. The second delta function enables us
to replace $g_k$ by $\epsilon/t (-1)^\lambda$, yielding to
\begin{eqnarray}
\rho(\omega)&=&\sum_{\nu}\int\limits_{-\infty}^\infty d\epsilon
\delta[\omega-t'\frac{\epsilon^2}{t^2}+3t'-(-1)^{\nu}\epsilon]\nonumber \\
&\times& \frac {1}{2N}\sum_{k\lambda}\delta[\epsilon-(-1)^\lambda t
g_k]\nonumber \\
&=&\sum_{\nu}\int\limits_{-\infty}^\infty \frac{d\epsilon}{2}
\delta[\omega-t'\frac{\epsilon^2}{t^2}+3t'-(-1)^{\nu}\epsilon]\rho_0(\epsilon),\nonumber \\
\end{eqnarray}
where $\rho_0(\epsilon)$ is the DOS with $t'=0$, and then
\begin{eqnarray} \label{Dostp}
\rho (\omega )&=&{{\frac{1}{\sqrt{1+4\frac{t^{\prime
}}{t^{2}}(\omega
+3t^{\prime })}}}\frac{2}{\pi ^{2}}\sum_{\nu}{\frac{|\widetilde{\omega}_{\nu}|}{t^{2}}}\frac{1%
}{\sqrt{Z_{0}}}\mathbf{K}[Z(\widetilde{\omega}_{\nu})]}, \nonumber \\
\widetilde{\protect\omega}_{\nu}&=&{\frac{t^{2}}{2t^{^{\prime }}}}[1+(-1)^{\nu}\protect\sqrt{%
1+4\frac{t^{\prime }}{t^{2}}(\protect\omega +3t^{\prime })}].
\protect
\end{eqnarray}

Now let us turn to evaluate the NMR relaxation rate and the static
spin susceptibility. In general, $1/T_{1}$ measures the local
dynamics of the spins, and it is related to the transverse spin
susceptibility $\chi_{\perp}(i\omega_{n})$, which reads as\cite{ss}
\begin{equation}
\chi_{\perp}(i\omega_n)=-\int\limits_0^\beta\textmd{d}\tau
e^{i\omega_n\tau}\-\langle T_\tau
S_{i}^{\dag}(\tau)S_{i}^{-}(0)\rangle,
\end{equation}
and after a straightforward calculation, we obtain
\begin{eqnarray}
\langle T_\tau
S_{i}^{\dag}(\tau)S_{i}^{-}(0)\rangle=\sum_{m,n}g_{mn}(0,\tau)g_{mn}(0,-\tau).
\end{eqnarray}
With the help of spectral representation, we define
\begin{eqnarray}
A(k,\omega)=-2{\rm Im}\sum_{m,n}g_{mn} (k,\omega),
\end{eqnarray}
which has a relationship with the DOS
\begin{equation}
{1\over N}\sum_{k}A(k,\omega)=2\pi \rho(\omega).
\end{equation}
In this representation, we can obtain the NMR relaxation rate
finally
\begin{eqnarray}\label{st}
\frac{1}{T_{1}T}&=&\lim_{\omega\rightarrow0}\frac{k_{B}}{g^{2}\mu^{2}_{B}\hbar^{2}}\sum_{k}F^{2}(k)\frac{{\rm
Im}\chi_{\perp}(k,\omega)}{\omega} \nonumber \\
&=&\pi F^{2}\int^{\infty}_{-\infty}\rho^{2}(\omega')\frac{\beta
}{4}\frac{{d\omega'}}{\cosh^{2}(\beta \frac{\omega'-\mu}{2})},
\end{eqnarray}
in which $\mu_{B}$ is the Bohr magneton, $g$ is the electron $g$
factor, $\hbar$ is the Planck constant, and $\beta=1/k_{B}T$, where
$k_{B}$ is the Boltzman constant. The $F(k)$ are hyperfine form
factors, which do not vary much with $k$ in general. So we write all
these form factors as $F$, which is independent of temperature. In
the similar way, the static spin susceptibility can be derived as
\begin{eqnarray}\label{ss}
\chi&=&\lim_{q\rightarrow0}\lim_{\omega\rightarrow0}\chi(q,\omega) \nonumber \\
&=&g^{2}\mu^{2}_{B}\int^{\infty}_{-\infty}\rho(\omega')\frac{\beta
}{4}\frac{{d\omega'}}{\cosh^{2}(\beta \frac{\omega'-\mu}{2})},
\end{eqnarray}
and $\chi(q,\omega)$ comes from
\begin{equation}
\chi(i-j,\tau)=-\int\limits_0^\beta\textmd{d}\tau
e^{i\omega_n\tau}\-\langle T_\tau
S_{i}^{z}(\tau)S_{j}^{z}(0)\rangle.
\end{equation}

\begin{figure}
\includegraphics[scale=0.55]{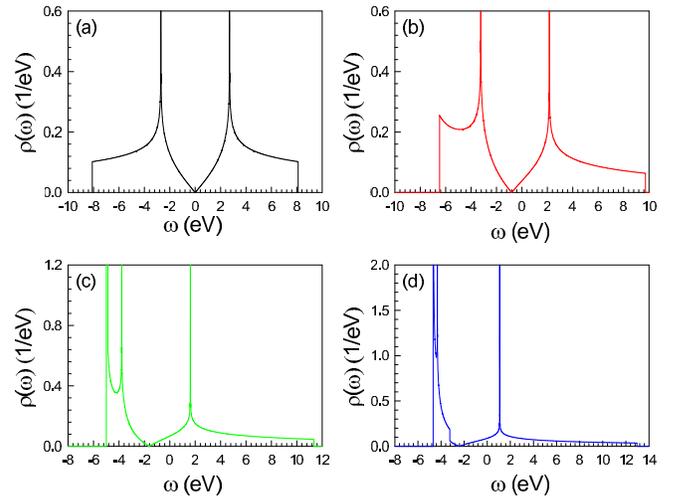}
\caption{DOS for different values of the next nearest neighbor
hopping $t'$:(a) $t'$=0; (b) $t'$=0.27 eV; (c) $t'$=0.54 eV; (d)
$t'$=0.81 eV with $t$=2.7 eV. }
\end{figure}

\section{Results and discussions}
NMR is a powerful method to characterize correlated states of
materials as it is sensitive to the DOS near the Fermi edge. So we
study the DOS firstly, and our results for different values of the
next neighbor hoping $t'$ have been shown in Fig.1. Focusing on the
particle-hole symmetric case, in Fig.1 (a), it is clear that,
besides the linear vanishing of the DOS at the Fermi level, there
are marked van~Hove singularities at the hopping energy, $\omega=\pm
t$. With a finite $t'$, as shown in Fig.1 (b), (c) and (d), these
van~Hove singularities shall appear at $\widetilde{\omega}_{\nu}=\pm
t$, which is derived from Eq. 12, and we will show that $t'$ plays
an important role in graphene as it breaks the particle-hole
symmetry.

For both $t'=0$ and $t'\neq0$ cases, the DOS in graphene is markedly
different from that in normal metals\cite{Dos} as its low energy
excitations are two-dimensional massless Dirac
fermions\cite{Semenof,Geim}, and the presence of $t'$ shifts in
energy the position of the Dirac point and breaks particle-hole
symmetry. Our further results shall indicate that these kinds of
important features of the DOS are at the origin of interesting
properties of the NMR relaxation rate and the static spin
susceptibility in graphene, as well as many transport anomalies in
this material\cite{Rmp}.

Having been familiar with the main features of the DOS in graphene,
we now turn to the evaluation of the NMR relaxation rate and the
static spin susceptibility. In Fig.2 (a) and (b), the NMR relaxation
rate and the static spin susceptibility as a function of $t'$ for
$t$=2.7 eV, $T$=10K at $\mu$=0 are plotted respectively. As shown by
the dark line in Fig.2 (a), the NMR relaxation rate follows a
$t'^{2}$ power law at $\mu=0$, while the static spin susceptibility
is linearly dependent on $t'$ as indicated by the dark line in
Fig.2(b). One of the most important properties of the NMR relaxation
rate is its temperature dependent behavior. In the inset of Fig.2(a)
and (b), we have plotted the temperature dependence of the NMR
relaxation rate and the static spin susceptibility respectively at
$t$=2.7 eV, $t'$=0 and $\mu$=0. As indicated by the dark lines with
triangle, the NMR relaxation rate follows a $T^{2}$ power law, and
the static spin susceptibility is linearly dependent on temperature
$T$.

\begin{figure}
\includegraphics[scale=0.5]{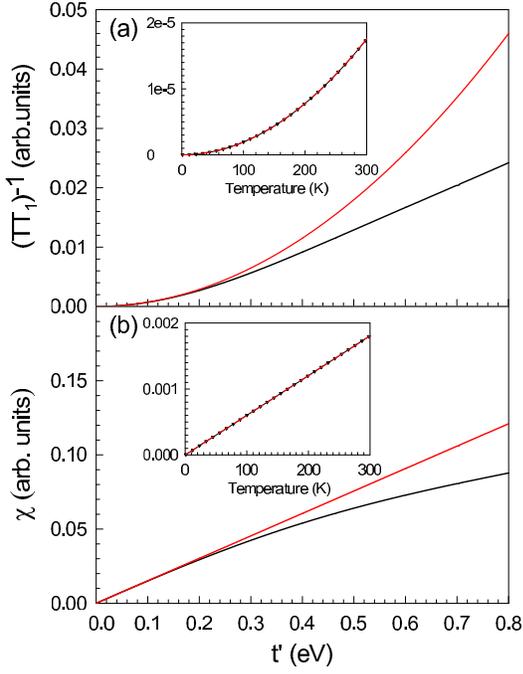}
\caption{(a) The NMR relaxation rate and (b) the static spin
susceptibility as function of $t'$ for $t$=2.7 eV, $T$=10K at
$\mu$=0. Dark lines indicate the exact numerical result and red
lines indicate data computed from Eq. \ref{sta} or Eq. \ref{ssa}.
Inset: (a) The NMR relaxation rate and (b) the static spin
susceptibility as function of temperature $T$ for $t$=2.7 eV,
$t'$=0, and $\mu$=0. Dark lines with triangle indicate the exact
numerical result and dash red lines indicate data computed from Eq.
\ref{sta} or Eq. \ref{ssa}. }
\end{figure}

\begin{figure}
\includegraphics[scale=0.5]{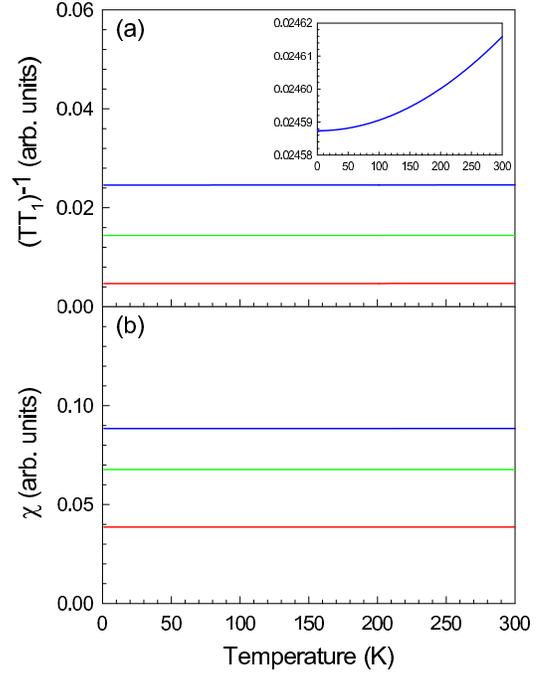}
\caption{(Color online )(a) The NMR relaxation rate and (b) the
static spin susceptibility as function of temperature for $t$=2.7 eV
with $t'$=0.27 eV (red line), $t'$=0.54 eV (green line), $t'$=0.81
eV (blue line) at $\mu$=0. Inset: Enlarge scale for $t'$=0.81 eV. }
\end{figure}

With a finite chemical potential $\mu$ and $t'$, the approximate
behavior of the NMR relaxation rate and the static spin
susceptibility may be expressed analytically at low temperatures.
From Eq.\ref{st}, we see that only low energy part shall contribute
to the NMR relaxation rate due to the properties of Fermi function
at low temperatures. When $|\widetilde{\omega}_{\nu}|\ll t$, as well
as $t'\ll t$, the DOS may be expressed as
\begin{equation}\label{Dos}
\rho(\omega)\simeq\frac{2\sqrt{3}}{3\pi} \frac{|\omega+3t'|}{t^2}
\end{equation}
approximately, and finally
\begin{eqnarray}\label{sta}
\frac{1}{T_{1}T}\simeq F^{2} \frac{4}{3\pi}
\frac{1}{t^4}[(\mu+3t')^{2}+\frac{\pi^{2}}{3}k^{2}_{B}T^{2}].
\end{eqnarray}
In this similar way, the static spin susceptibility may be
approximated by
\begin{eqnarray}\label{ssa}
\chi\simeq g^{2}\mu _{B}^{2}\frac{2\sqrt{3}}{3\pi
}\frac{1}{t^{2}}\{(\mu+3t')+2k_{B}T\ln [1+e^{-\beta (\mu+3t')}]\}.
\end{eqnarray}

The data computed within Eq.\ref{sta} and Eq.\ref{ssa} have also
been shown in Fig.2, which are indicated by red lines, and each of
them is very near to the exact result especially when $t'<0.2$eV.

The calculated temperature dependence of the NMR relaxation rate and
the static spin susceptibility for different values of $t'$ at
$\mu=0$ are shown in Fig.3 (a) and (b) respectively. At first
glance, it seems that the NMR relaxation rate and the static spin
susceptibility are almost independent of temperature for a fixed
$t'$. In the inset of Fig.3(b), we enlarged the scale for the case
of $t'$=0.81 eV, and the NMR relaxation rate increases as
temperature increases, following a $T^{2}$ power law. However, the
enhancement with temperature is rather small comparing with the
whole trend.

This unusual phenomenon is a direct results of the low energy
excitations of graphene, which behave as massless Dirac fermions. In
low energy regime, the DOS in graphene is linear around the
particle-hole symmetric filling, and vanishes at the Dirac point,
while the presence of $t'$ shifts in energy the position of the
Dirac point. Mathematically it is clearly seen through Eq.\ref{Dos}.
Hence, with a finite chemical potential $\mu$ and $t'$, we can
describe our result within Eq.\ref{sta} and Eq.\ref{ssa} at low
temperatures rather well. In the particle-hole symmetric case,
namely, $t'$=0, the NMR relaxation rate follows a $T^{2}$ power law.
However, the $T^{2}$ term in the NMR relaxation rate is negligible
with respect to $(\mu+3t')$ for realistic values of $\mu$ and $t'$.
On the static spin susceptibility, it is linearly dependent on $T$
at half filling when $t'=0$, while with a finite $\mu$ and $t'$, it
should be dominated by $(\mu+3t')$ terms in low energy regime, which
may be described by Eq.\ref{ssa} very well. Our results also show
that $t'$ plays an important role in graphene since it breaks the
particle-hole symmetry and is responsible for various effects
observed experimentally.

\begin{figure}
\includegraphics[scale=0.5]{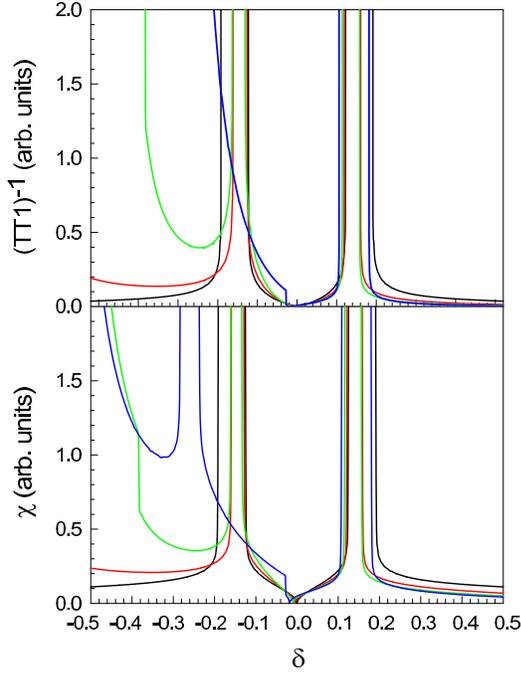}
\caption{(Color online) (a) The NMR relaxation rate and (b) the
static spin susceptibility as function of the excess charge for
$t$=2.7 eV with $t'$=0 (dark line), $t'$=0.27 eV (red line),
$t'$=0.54 eV (green line) and $t'$=0.81 eV (blue line) at T=10K. }
\end{figure}

Arguably, one of the most interesting and promising properties from
the technological point of view is the ability to tune the carrier
density in graphene through a gate voltage\cite{Novoselov}. Now
let's turn to study the case with a finite excess charge $\delta$,
and the chemical potential $\mu$ is determined by
\begin{eqnarray}\label{u}
\int^{\infty}_{-\infty}\rho(\omega)[\frac{1}{e^{\beta(\omega-\mu)}+1}-\frac{1}{2}]{d\omega}=\delta.
\end{eqnarray}
With the help of the gate voltage, one can control the density and
type ( $n$ or $p$ ) of carriers varying their chemical
potential\cite{Geim}. The calculated $\delta$ dependence of the NMR
relaxation rate and the static spin susceptibility for different
values of $t'$ have been shown in Fig.4 (a) and (b) respectively.
The NMR relaxation rate and the static spin susceptibility are
linearly dependent on $|\delta|$ when $|\delta|<0.1$ except
$t'$=0.81 eV. While $\delta$ is high enough, it is interesting to
find that there are several prominent peaks. In the particle-hole
symmetric case, peaks appear around the center at
$\delta$=$\pm$0.157. With a finite $t'$, these peaks shall appear
around the center at $\delta$=0.140, $-$0.144 for $t'$=0.27 eV,
$\delta$=0.136, $-$0.147 for $t'$=0.54 eV, and $\delta$=0.144,
$-$0.260 for $t'$=0.81 eV.

These intriguing phenomenon may be predicted from the behavior of
the DOS in graphene directly. When the chemical potential is located
at the marked van~Hove singularities, where the DOS peaks
dramatically, hence peaks shall appear in the NMR relaxation rate
and the static spin susceptibility\cite{Peak}. To learn more on
these peaks, the calculated $\delta$ dependence of the NMR
relaxation rate and the the static spin susceptibility with $t$=2.7
eV, $t'$=0.27 eV at different temperatures are plotted in Fig.5 (a)
and (b) respectively. These peaks in the NMR relaxation rate and the
static spin susceptibility decrease as the temperature increases,
however, these peaks are even pronounced at 300 K, and therefore,
there peaks should produce a distinct effect on temperature
dependence of the NMR relaxation rate and the static spin
susceptibility.
\begin{figure}
\includegraphics[scale=0.5]{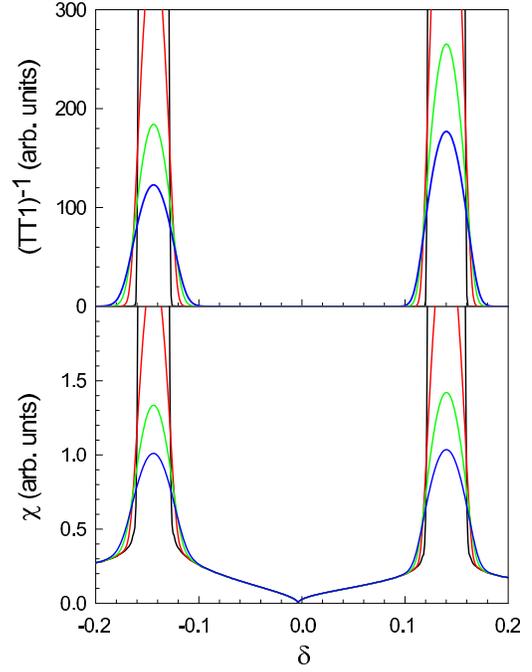}
\caption{(Color online) (a) The NMR relaxation rate and (b) the
static spin susceptibility as function of $\delta$ for $t$=2.7 eV,
$t'$=0.27 eV at different temperatures: $T$=10K (dark line),
$T$=100K (red line), $T$=200K (green line) and $T$=300K (blue line).
}
\end{figure}

In Fig.6 (a) and (b), we plot the NMR relaxation rate and the static
spin susceptibility as function of temperature at different $\delta$
for $t$=2.7 eV and $t'$=0.27 eV. Around $\delta$=0.12, we can
immediately separate the data into two temperature regimes by a
crossover: the high temperature regime and the low temperature
regime. The NMR relaxation rate and the static spin susceptibility
increase dramatically as temperature increases in the low
temperature regime, and after the crossover, both of them decrease
as temperature increases, and the position of temperature dependent
crossover is sensitive to $\delta$.

For a fixed $\delta$, the chemical potential varies slightly as
temperatures varies, so when the chemical potential associated with
the temperature is located at the van~Hove singularities, where the
DOS peaks dramatically, there shall be peaks in both of the NMR
relaxation rate and the static spin susceptibility, which may divide
the temperature dependence of the NMR relaxation rate and the static
spin susceptibility into two temperature regimes. The temperature at
the crossover decreases as $\delta$ increases when $\delta<0.140$.
As $\delta$ tends toward $0.140$, the NMR relaxation rate and the
static spin susceptibility shall decrease as temperature increases
almost in the whole temperature regime for current parameters, for
instance, pink lines in Fig.6 (a) and (b); and afterward, the
temperature at the crossover increases as $\delta$ increases when
$\delta>0.140$ (dash lines in Fig. 6 (a) and (b) are respect to this
illumination ). We only analyze the case when $\delta>0$ at
$t'=0.27$ eV; because, for other values of $t'$ or when $\delta<0$,
the essential nature of peak's appearing is the same.

\begin{figure}
\includegraphics[scale=0.5]{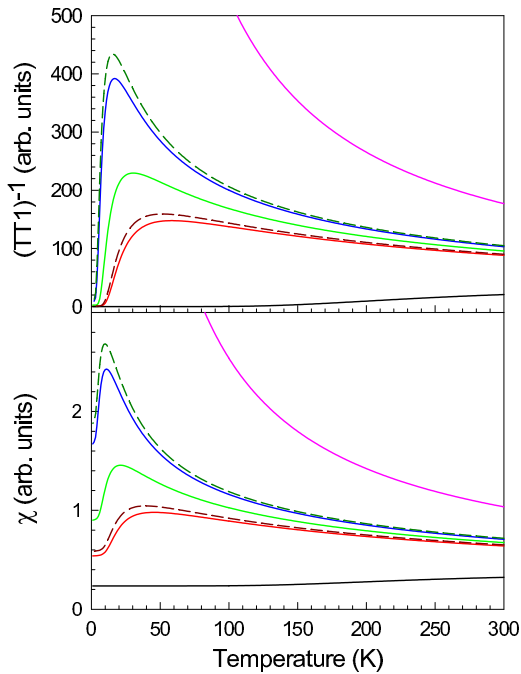}
\caption{(Color online) (a) The NMR relaxation rate and (b) the
static spin susceptibility as function of temperature $T$ for
$\delta$=0.11 (dark line), $\delta$=0.12 (red line), $\delta$=0.121
(green line), and $\delta$=0.122 (blue line), $\delta$=0.14 (pink
line), $\delta$=0.158 (dash-dark-green line), and $\delta$=0.16
(dash-dark-red line) at $t$=2.7 eV and $t'$=0.27 eV. }
\end{figure}

\section{Summary}

In summary, we have analyzed the NMR relaxation rate and the static
spin susceptibility in graphene, which show a behavior that is not
that of a normal metal. In low energy regime, the DOS in graphene is
linear around the particle-hole symmetric filling, and vanishes at
the Dirac point, while the presence of $t'$ shifts in energy the
position of the Dirac point. Hence, the NMR relaxation rate follows
a power law as $T^2$ on the particle-hole symmetric side at half
filling, while away from half filling and with a finite $t'$, the
$(\mu+3t')^2$ terms dominate at low excess charge. The static spin
susceptibility is linearly dependent on $T$ at half filling when
$t'=0$, while with a finite $\mu$ and $t'$, it should be dominated
by $(\mu+3t')$ terms in low energy regime. The next-nearest neighbor
$t'$ plays an important role in graphene as it breaks the
particle-hole symmetry and is responsible for various effects
observed experimentally. These unusual phenomena are direct results
of the low energy excitations of graphene.

The NMR relaxation rate and the static spin susceptibility are
linearly dependent on the excess charge $|\delta|$ when $|\delta|$
is small, while at high $\delta$, there is a pronounced crossover
which divides the temperature dependence of the NMR relaxation rate
and the static spin susceptibility into two temperature regimes: the
NMR relaxation rate and the static spin susceptibility increase
dramatically as temperature increases in the low temperature regime,
and after the crossover, both decrease as temperature increases at
high temperatures. This crossover is due to the well-known
logarithmic Van Hove singularity in the DOS, and its position
dependence of temperature is sensitive to $\delta$. These properties
show that graphene is a new class of materials with an unusual
metallic state. Since the electronic density is easily controlled by
a gate voltage, these phenomena can certainly be tested
experimentally.

\begin{acknowledgments}
We acknowledge useful discussions with P. Thalmeier and K. Ziegler.
This work was supported by the Hungarian Scientific Research Fund
under grant number OTKA K72613.

\end{acknowledgments}


\clearpage

\end{document}